\documentclass[12pt,a4paper,final]{iopart}
\usepackage{iopams}
\usepackage{float}
\usepackage{graphicx}
\usepackage[breaklinks=true,colorlinks=true,linkcolor=blue,urlcolor=blue,citecolor=blue]{hyperref}
\usepackage{tabularx}
\makeatletter
\newcommand{\rmnum}[1]{\romannumeral #1}
\newcommand{\Rmnum}[1]{\expandafter\@slowromancap\romannumeral #1@}
\makeatother

\begin{document}

\title{``Hard" crystalline lattice in the Weyl semimetal NbAs}

\author{Yongkang Luo$^{1*}$, N. J. Ghimire$^1$\footnote[1]{Present address: Materials Science Division, Argonne National Laboratory, Argonne, Illinois 60439, USA}, E. D. Bauer$^1$, J. D. Thompson$^1$, and F. Ronning$^{1\ddag}$}

\address{$^{1}$ Los Alamos National Laboratory, Los Alamos, New Mexico 87545, USA}

\ead{ykluo@lanl.gov; fronning@lanl.gov}

\date{\today}

\begin{abstract}
We report the effect of hydrostatic pressure on the magnetotransport properties of the Weyl semimetal NbAs. Subtle changes can be seen in the $\rho_{xx}(T)$ profiles with pressure up to 2.31 GPa. The Fermi surfaces undergo an anisotropic evolution under pressure: the extremal areas slightly increase in the $\mathbf{k_x}$-$\mathbf{k_y}$ plane, but decrease in the $\mathbf{k_z}$-$\mathbf{k_y}$($\mathbf{k_x}$) plane. The topological features of the two pockets observed at atmospheric pressure, however, remain unchanged at 2.31 GPa. No superconductivity can be seen down to 0.3 K for all the pressures measured. By fitting the temperature dependence of specific heat to the Debye model, we obtain a small Sommerfeld coefficient $\gamma_0=$ 0.09(1) mJ/(mol$\cdot$K$^2$) and a large Debye temperature, $\Theta_D=$ 450(9) K, confirming a ``hard" crystalline lattice that is stable under pressure. We also studied the Kadowaki-Woods ratio of this low-carrier-density massless system, $R_{KW}=$ 3.2$\times 10^4$ $\mu\Omega$ cm mol$^2$ K$^2$ J$^{-2}$. After accounting for the small carrier density in NbAs, this $R_{KW}$ indicates a suppressed transport scattering rate relative to other metals. \\

\hspace{-15pt}\textbf{Keywords:} Weyl semimetal, magnetotransport, Shubnikov-de Hass oscillation, Kadowaki-Woods ratio

\end{abstract}

\pacs{75.47.-m, 61.50.Ks, 71.70.Di, 73.20.At}

\maketitle

\section{Introduction}

Recently, the Weyl fermion, a long-sought quasiparticle in high-energy physics\cite{Weyl1929}, has been discovered in condensed-matter physics in the so-called Weyl semimetals (WSMs). WSMs are related to Dirac semimetals (DSMs). The low-energy excitations in a DSM are described by Dirac points in the Brillouin zone where conduction- and valence-bands linearly cross and form a relativistic dispersion. Such massless Dirac fermions are protected by both time reversal symmetry and crystalline inversion symmetry. Once either of them is broken, the four fold degenerate Dirac point splits into two Weyl points with opposite chiralities. These Weyl points act as the sources or drains of Chern flux, {\it viz.} magnetic monopoles in momentum space\cite{Balents-Weyl,Hosur-WSM,Wan-5dSpinel,Wang-WSM2013}. Semimetals with Weyl points in band structure are called WSMs. Like its ``parent" compound DSM, a WSM manifests itself with many exotic properties. One way to find a WSM is to look into the materials that lack spatial inversion symmetry, and this has been realized in a class of binary transition-metal monopnictides of the form $TmPn$ (where $Tm$ = Ta or Nb, and $Pn$ = As or P)\cite{Weng-TmPn}. The intriguing Weyl nodes in the bulk and the Fermi-arc surface states have been observed by angle-resolved photoemission spectroscopy (ARPES) experiments\cite{XuS-TaAsARPES,Lv-TaAsPRX,Lv-TaAsNP,XuS-NbAsARPES}, which are in good agreement with theoretical predictions\cite{Hosur-Friedel,Ojanen-WSMFmArc,Potter-QOFmArc}. Magnetotransport measurements also revealed large unsaturated magnetoresistances, low carrier densities, ultrahigh mobilities, non-trivial Berry phases and Adler-Bell-Jackiw chiral anomalies \cite{Zhang-TaAsSdH,Huang-TaAsLMR,Shekhar-NbP,Nirmal-NbAs,WangZ-NbPSdH,LuoY-NbAsSdH,YangX-NbAsLMR,Shekhar-TaPLMR,Son-ChirAnom}, signaling a promising potential in device applications.

Pressure is an effective non-thermal control parameter. In general, pressure tunes the properties of a material by three means. (\rmnum{1}) By compacting the crystalline structure, pressure broadens the electronic bands and enhances the intersite hopping rate (mobility); (\rmnum{2}) It affects the extent of electronic correlations, typically in Kondo lattice compounds in which a variety of ground states including magnetism, non-Fermi liquid, quantum criticality, unconventional superconductivity (SC) and heavy fermion (HF) liquid may appear in different pressure regions\cite{Doniach}. (\rmnum{3}) More interestingly, by changing the crystalline structure, the properties of a material could be significantly changed. One particularly relevant example is Cd$_3$As$_2$, a representative DSM\cite{Wang-Cd3As2DSM}. Cd$_3$As$_2$ undergoes a structural phase transition from a tetragonal phase in space group I4$_1$/acd to a monoclinic phase in space group P2$_1$/c at a critical pressure 2.57 GPa, above which the DSM state breaks down and the system enters a gapped semiconducting state\cite{ZhangS-Cd3As2Pre}; SC emerges out of this semiconducting phase when pressure exceeds 8.5 GPa\cite{HeL-Cd3As2SC,WangH-Cd3As2SC}.

There have only been a few studies on the effect of pressure on the members of $TmPn$ family of WSMs\cite{Zhang-NbAsPre,ZhouY-TaAsPre}. Herein, we investigated the effect of hydrostatic pressure on the magnetotransport properties of \emph{high-quality} NbAs single crystals. We find that the $\rho_{xx}(T)$ profile is almost unchanged with pressure up to 2.31 GPa, except for a small enhancement of residual resistivity at low temperature. Shubnikov-de Hass (SdH) oscillation measurements reveal that the Fermi surfaces undergo an anisotropic evolution under pressure: the extremal areas slightly increase in the $\mathbf{k_x}$-$\mathbf{k_y}$ plane, but decrease in the $\mathbf{k_z}$-$\mathbf{k_y}$($\mathbf{k_x}$) plane. The topological features of the two observed pockets remain unvaried. No SC can be seen down to 0.3 K for all the pressures measured.
By fitting the temperature dependence of specific heat to the Debye model, we obtain a large Debye temperature for this compound, $\Theta_D=$ 450(9) K, confirming a ``hard" crystalline lattice that is stable under pressure. The Kadowaki-Woods ratio of this low-carrier-density massless system is also studied.

\section{Experimental}

Millimeter sized single crystals of NbAs were synthesized by a vapor transport method with iodine as described elsewhere\cite{Nirmal-NbAs}. The crystallographic orientation of the sample was verified by X-ray diffraction (XRD). In-plane electrical transport measurements ($\textbf{I}\parallel\textbf{a}$) were made on two samples (labeled by S1 and S2) as functions of pressure and magnetic field. The two samples were cut from the same piece of a plate-like single crystal. S1 was mounted with $\textbf{B}\parallel\textbf{c}$, while the measurements on S2 were in the configuration of $\textbf{B}\parallel\textbf{a}$. A piston-clamp type pressure cell with Daphne 7373 oil as the pressure medium, was used to generate hydrostatic pressures up to 2.31 GPa, and pressure was determined by measuring the superconducting transition of Pb. Both longitudinal resistivity ($\rho_{xx}$) and transverse (Hall) resistivity ($\rho_{yx}$, S1 only) were measured by slowly sweeping a DC magnetic field from $-$9 T to 9 T at a rate of 0.2 T/min. An AC-resistance bridge (LR-700) was employed to perform these transport measurements in a Heliox refrigerator. Specific heat measurements were carried out using a commercial calorimetry option on a Physical Property Measurement System (PPMS-9, Quantum Design). Three single crystals with a total mass $\sim$~24 mg were accumulated on the puck. Specific heat for 0.4-300 K was measured by a thermal relaxation method.

\section{Results and Discussion}

\begin{figure}[htbp]
\includegraphics[width=9.5cm]{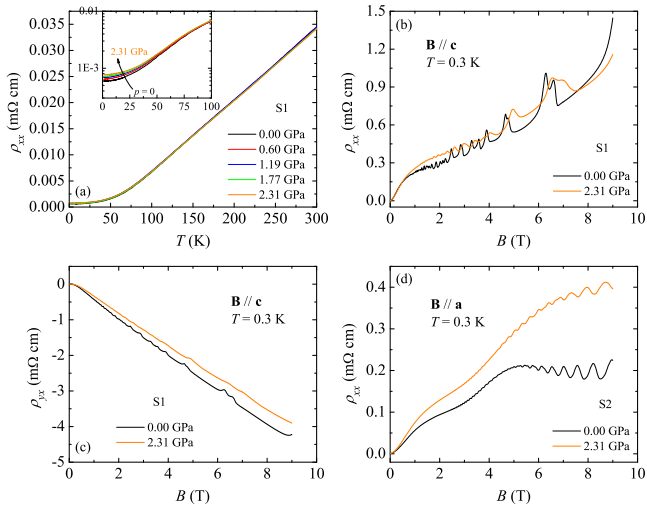}
\caption{\label{Fig.1} (a) Temperature dependence of $\rho_{xx}$ under various pressures. The inset shows the low temperature part plotted on a semi-$\log$ scale. (b) and (c) respectively show field dependent $\rho_{xx}$ and $\rho_{yx}$ at $T=$ 0.3 K in the configuration of $\textbf{B}\parallel\textbf{c}$. (d) displays $\rho_{xx}(B)$ with $\textbf{B}\parallel\textbf{a}$.}
\end{figure}

Fig.~\ref{Fig.1} shows the temperature dependence of resistivity at various pressures. As mentioned in our earlier reports\cite{Nirmal-NbAs}, NbAs behaves as a good metal in the absence of magnetic field. $\rho_{xx}(T)$ is almost a linear function of $T$ before it starts to saturate and show Fermi-liquid-like behavior at low temperature [inset to Fig.~\ref{Fig.4}(c)]. The measured S1 sample possesses a residual resistance ratio $RRR$[$\equiv$$\rho_{xx}(300 K)/\rho_{xx}(0.3 K)$]$\simeq$60, comparable with previous results\cite{Nirmal-NbAs,YangX-NbAsLMR}. Somewhat surprisingly, there is negligible change in $\rho_{xx}(T)$ for pressures up to 2.31 GPa: all the $\rho_{xx}(T)$ curves almost collapse onto a single line, and the difference among them is approaching the precision of the measurements. Tiny changes of $\rho_{xx}(T)$ can only be seen at low temperatures, in a $\log$ scale in the inset to Fig.~\ref{Fig.1}(a). Down to the lowest temperature we have reached, 0.3 K, there is no signature of superconductivity for all the pressures measured. At 0.3 K, the residual resistivity is 5.82$\times$10$^{-4}$ m$\Omega$ cm at atmospheric pressure, and increases to 7.66$\times$10$^{-4}$ m$\Omega$ cm at 2.31 GPa. It should be pointed out that our $\rho_{xx}(T)$ results are very different from Zhang {\it et al.}\cite{Zhang-NbAsPre} in which the authors observed a mysterious upturn at low temperature when pressure exceeds 1.5 GPa and the magnitude of resistivity significantly increases with pressure as well. Whether this difference depends on the sample quality needs to be confirmed in the future.

To better clarify this weak pressure effect on NbAs, we also performed SdH quantum oscillation measurements at $p$ = 0 and 2.31 GPa, and the data for 0.3 K are displayed in Figs.~\ref{Fig.1}(b-d). In the case of $\textbf{B}\parallel\textbf{c}$, pronounced SdH oscillations are observed in both $\rho_{xx}(B)$ and $\rho_{yx}(B)$. Oscillatory signals are also seen in the configuration $\textbf{B}\parallel\textbf{a}$. One important feature for the data at $p$ = 2.31 GPa is that the amplitude of the SdH oscillations becomes smaller when compared to the atmospheric curves, in both $\textbf{B}\parallel\textbf{c}$ and $\textbf{B}\parallel\textbf{a}$ configurations.

\begin{figure*}[htbp]
\hspace*{-35pt}
\includegraphics[width=18.5cm]{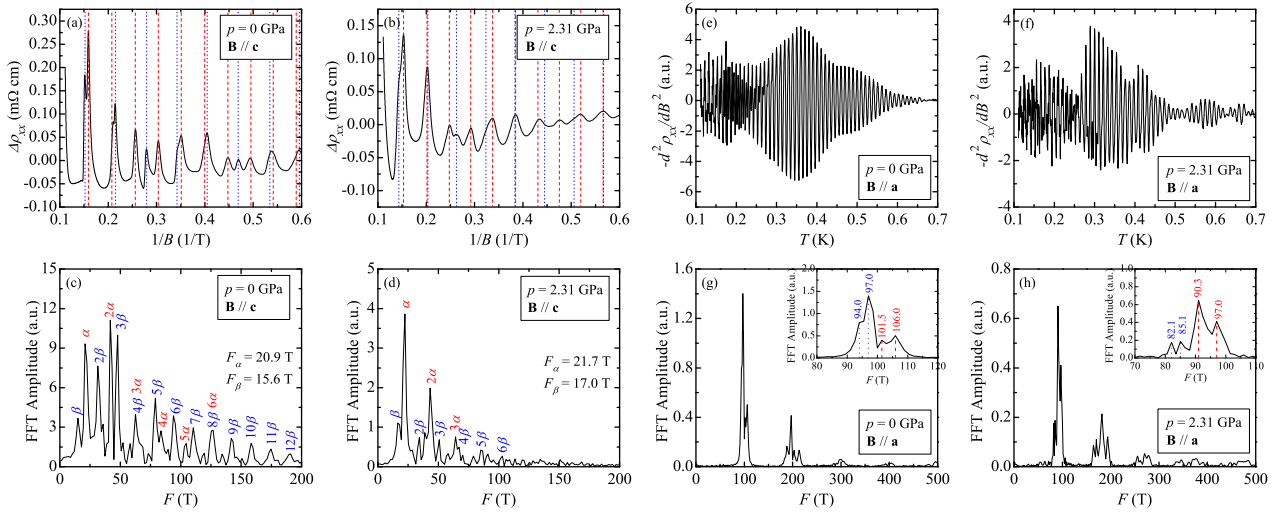}
\caption{\label{Fig.2} SdH oscillations of NbAs at $p=$ 0 and 2.31 GPa in the configurations of $\textbf{B}\parallel\textbf{c}$ (a-d) and $\textbf{B}\parallel\textbf{a}$ (e-h). All the data were taken at $T=$ 0.3 K. (a) and (b) show $\Delta\rho_{xx}$ as a function of $1/B$, while (e) and (f) are plots of $-d^2\rho_{xx}/dB^2$ vs. $1/B$. The lower panels are the FFT spectra of $-d^2\rho_{xx}/dB^2$. The insets to (g) and (h) are the zoom-in views.}
\end{figure*}

In order to analyze the  effect of pressure on the SdH oscillations of $\rho_{xx}$ in NbAs, we use the following expression for a three-dimensional (3D) system\cite{Zhang-TaAsSdH,Shoenberg,Murakawa-Rashba,Xiang-P}:
\begin{equation}
\frac{\Delta\rho_{xx}}{\langle\rho_{xx}\rangle}=A(T,B)\cos[2\pi(\frac{F}{B}-\gamma+\delta)],
\label{Eq.1}
\end{equation}
in which $\langle\rho_{xx}\rangle$ is the non-oscillatory part of $\rho_{xx}$, $F$ is the frequency of oscillation, $\gamma$ is the Onsager phase, and $\delta$ is an additional phase shift taking a value between $\pm1/8$ depending on the curvature of the Fermi surface (FS) topology\cite{Murakawa-Rashba,WangJ-Cd3As2SdH}. Fig.~\ref{Fig.2}(a) shows $\Delta\rho_{xx}$, the oscillating part of $\rho_{xx}$, as a function of $1/B$ at ambient pressure with $\textbf{B}\parallel\textbf{c}$. The data well reproduce our previous results\cite{LuoY-NbAsSdH}, and two sets of oscillatory frequencies can be clearly assigned; this is better seen in the Fast Fourier Transform (FFT) spectrum of $-d^2\rho_{xx}/dB^2$ as shown in Fig.~\ref{Fig.2}(c). The two frequencies are labeled as $F^c_{\alpha}$ = 20.9 T (The superscripts $c$ or $a$ denote the field orientation; the same hereafter) and $F^c_{\beta}$ = 15.6 T. From the Lifshitz-Onsager relation that the quantum oscillation frequency $F$ is correlated with the extremal cross-sectional area $S_F$ of the Fermi surface by $F=\frac{\hbar}{2\pi e}S_F$, the magnitude of Fermi wave vector for $\alpha$- and $\beta$-pockets are then calculated via $k_F=(S_F/\pi)^{1/2}$. The calculated values are $k_F^{\alpha c}=0.0252$ \AA$^{-1}$ and $k_F^{\beta c}=0.0218$ \AA$^{-1}$, respectively. According to our previous analysis combining with Hall effect measurements\cite{LuoY-NbAsSdH}, the $\alpha$-pocket is probably electron-type, but the $\beta$-pocket is hole-type. We should also point out that besides $F^c_{\alpha}$ and $F^c_{\beta}$, a series of their higher-order harmonics are seen in Fig.~\ref{Fig.2}(c), manifesting the high quality of the sample studied. At 2.31 GPa, both of the pockets slightly enlarge, and the frequencies become $F^c_{\alpha}$ = 21.7 T and $F^c_{\beta}$ = 17.0 T, see Fig.~\ref{Fig.2}(d). In addition, increasing pressure substantially reduces the number of higher-order harmonics.

For the case of $\textbf{B}\parallel\textbf{a}$ [see Fig.~\ref{Fig.2}(e-h)], four fundamental frequencies are observed at $p=$ 0, {\it i.e.} $F^a_{\alpha}=$ 101.5 T, $F^{a'}_{\alpha}=$ 106.0 T, $F^a_{\beta}=$ 94.0 T and $F^{a'}_{\beta}=$ 97.0 T. Our earlier work has revealed that both $\alpha$- and $\beta$-pockets are upright banana-like pockets\cite{LuoY-NbAsSdH}, sitting within the mirror planes $M_x$ and $M_y$\cite{Weng-TmPn}. Therefore, each $\alpha$- and $\beta$-type orbits of the FS has two kinds of projections on the $\mathbf{k_y}$-$\mathbf{k_z}$ plane, which gives rise to the four observed frequencies in total\cite{LuoY-NbAsSdH}. Under a pressure of 2.31 GPa, all four frequencies become smaller, with the values of $F^a_{\alpha}=$ 90.3 T, $F^{a'}_{\alpha}=$ 97.0 T, $F^a_{\beta}=$ 82.1 T and $F^{a'}_{\beta}=$ 85.1 T [cf the insets to Figs.~\ref{Fig.2}(g) and (h)]. The fact that the FS extremal area expands in the $\mathbf{k_x}$-$\mathbf{k_y}$ projection, but shrinks in both $\mathbf{k_y}$-$\mathbf{k_z}$ and $\mathbf{k_x}$-$\mathbf{k_z}$ projections manifests an anisotropic pressure effect.

\begin{figure}[htbp]
\includegraphics[width=8.5cm]{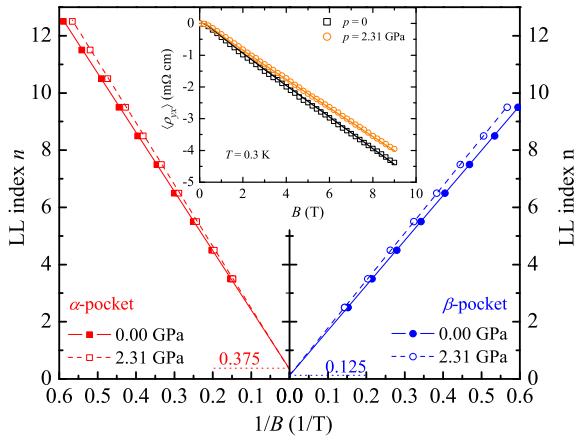}
\caption{\label{Fig.3} Landau level indices $n$ as functions of $1/B$ for $\textbf{B}\parallel\textbf{c}$. The solid and open symbols represent the data points at $p=$ 0 and 2.31 GPa, respectively. The two dot lines are guidelines for 0.125 and 0.375, respectively. The inset displays a comparison of $\langle\rho_{yx}(B)\rangle$ at 0.3 K between ambient pressure and 2.31 GPa. The lines through the symbols are two band fits to Eq.~(\ref{Eq.3}).}
\end{figure}

\begin{figure}[htbp]
\includegraphics[width=8cm]{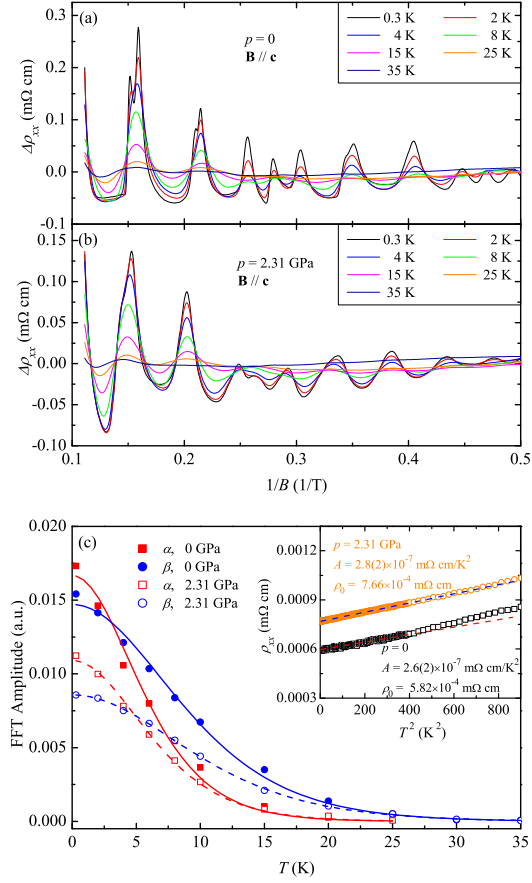}
\caption{\label{Fig.4} (a) and (b) respectively display SdH oscillations at selected temperatures under $p=$ 0 and 2.31 GPa, where $\Delta\rho_{xx}=\rho_{xx}-\langle\rho_{xx}\rangle$. (c) The ampiltudes of FFT on $\Delta\rho_{xx}$ for $\alpha$- and $\beta$-pockets. The lines are numerical fits to the LK formula. Inset, $\rho_{xx}$ versus $T^2$. }
\end{figure}

The Onsager phase is $\gamma=1/2-\Phi_B/2\pi$ with $\Phi_B$ being a Berry phase. In a normal metal, $\Phi_B=$ 0, and therefore $\gamma=$ 1/2. Whereas in a topological material, when a particle completes a closed trajectory that encloses a degenerate point with linear band dispersion, it acquires a geometrical phase $\Phi_B=~\pi$\cite{Berry-1984,Zak-PhiB}. This quantum correction leads to $\gamma=$ 0 around a Weyl point. Taking into account the additional phase $\delta$, we may expect that $\gamma-\delta$ is in the vicinity of 0 for a non-trivial $\pi$ Berry phase, but falls between 0.375 and 0.625 for a trivial Berry phase of 0. Previous studies of the SdH effect have uncovered that the Berry phase is 0 for the $\alpha$-pocket, and is $\pi$ for the $\beta$-pocket, suggesting that the $\alpha$-pocket is topologically trivial while the $\beta$-pocket is not\cite{LuoY-NbAsSdH}. We find that this feature is unchanged with pressure up to 2.31 GPa. In Fig.~\ref{Fig.3} we show  a Landau-fan diagram for $\textbf{B}\parallel\textbf{c}$. Here, we still follow our previous definition that the peak position of $\Delta\rho_{xx}$ is assigned as the half-integer ($n+1/2$) Landau level (LL) index\cite{LuoY-NbAsSdH}. By extrapolating the LL index to the infinite field limit, the intercepts (=$\gamma-\delta$) can be derived, and the corresponding Berry phases are also inferred as shown in Table \ref{Tab.1}. The results demonstrate that no topological phase transition takes place up to 2.31 GPa.

We also measured the SdH effect at various temperatures and at $p=$ 0 and 2.31 GPa, data of which are displayed in Figs.~\ref{Fig.4}(a) and (b), respectively. The decay of the SdH oscillation amplitude with increasing temperature conforms to the Lifshitz-Kosevich (LK) formula\cite{Zhang-TaAsSdH,Murakawa-Rashba}
\begin{equation}
A(T,B)\propto\exp(-2\pi^2k_BT_D/\hbar\omega_c)\frac{2\pi^2k_BT/\hbar\omega_c}{\sinh(2\pi^2k_BT/\hbar\omega_c)},
\label{Eq.2}
\end{equation}
in which $T_D$ is the Dingle temperature, and $\omega_c=\frac{eB}{m^*}$ is the cyclotron frequency with $m^*$ being the effective mass. Since many higher-order harmonics are observed in our sample [Figs.~\ref{Fig.2}(c) and (d)], and each of them has different masses but will contribute to the same peaks in $\Delta\rho_{xx}$/$\langle\rho_{xx}\rangle$, we hence fit the temperature dependent amplitude of the FFT on $\Delta\rho_{xx}$ to the LK formula, see Fig.~\ref{Fig.4}(c). The resultant effective masses at ambient pressure are $m_{\alpha}^*=0.046~m_0$ and $m_{\beta}^*=0.029~m_0$ for $\alpha$- and $\beta$-pockets, respectively. These values are a little smaller than in our earlier work\cite{LuoY-NbAsSdH}. Application of a 2.31-GPa pressure slightly decreases the effective masses to $m_{\alpha}^*=0.041~m_0$ and $m_{\beta}^*=0.026~m_0$. More physical parameters are calculated and summarized in Table \ref{Tab.1}. Note, the magnitude of the Fermi energy $\varepsilon_F$ substantially increases with pressure, in both the electron-type $\alpha$-band and the hole-type $\beta$-band, implying that pressure plays a band-broadening effect in this material, at least in the $\boldsymbol{\Gamma}$-$\mathbf{X}$($\mathbf{Y}$) direction of the Brillouin zone.

To obtain the carrier density and mobility as a function of pressure, we fit $\langle\rho_{yx}(B)\rangle$ at 0.3 K to a two-band model\cite{LuoY-WTe2Hall}:
\begin{equation}
\rho_{yx}(B)=\frac{B}{e}\frac{(n_h\mu_h^2-n_e\mu_e^2)+(n_h-n_e)\mu_e^2\mu_h^2B^2}{(n_h\mu_h+n_e\mu_e)^2+[(n_h-n_e)\mu_e\mu_hB]^2},
\label{Eq.3}
\end{equation}
where $n$ and $\mu$ are respectively carrier density and mobility, and the subscript $e$ (or $h$) denotes electron (or hole). The results for $p=$ 0 and 2.31 GPa are displayed in the inset to Fig.~\ref{Fig.3}. For $p=$ 0, we get $n_e=10.7 \times10^{18}$ cm$^{-3}$, $n_h=9.4\times10^{18}$ cm$^{-3}$, $\mu_e$ = 1.9$\times$10$^{5}$ cm$^2$/Vs, and $\mu_h$ = 1.9$\times$10$^{6}$ cm$^2$/Vs. (\rmnum{1}), Both $n_e$ and $n_h$ decrease with pressure (see Table~\ref{Tab.1}), consistent with the fact that the Fermi volumes [$\propto (F^aF^{a'}F^c)^{0.5}$] of both $\alpha$- and $\beta$-pockets decrease. (\rmnum{2}), $\mu_e$ and $\mu_h$ also decrease with pressure (see also in Table \ref{Tab.1}), for which at first glance one may be surprised considering the band-broadening effect that pressure has on this compound. Note that the Fermi velocity ($v_F=\frac{\hbar k_F}{m^*}$) increases with pressure. One possibility to reconcile this inconsistency is that the scattering rate increases (or in other words, the transport lifetime $\tau_{tr}=\frac{m^*\mu}{e}$ and mean-free path $l_{tr}=v_F\tau_{tr}$ decrease) with pressure. This picture is also supported by the facts that the residual resistivity increases [Fig.~\ref{Fig.1}(a)] and that the amplitudes of the SdH oscillations [Figs.~\ref{Fig.1}(b-d)] and the higher-order harmonics contributions [Figs.~\ref{Fig.2}(c-d)] become weaker under pressure. Considering the anisotropic change of Fermi surface size, a strain effect due to a slightly inhomogeneous pressure may provide an explanation for all these observations.

\begin{figure}[htbp]
\vspace*{-15pt}
\includegraphics[width=8cm]{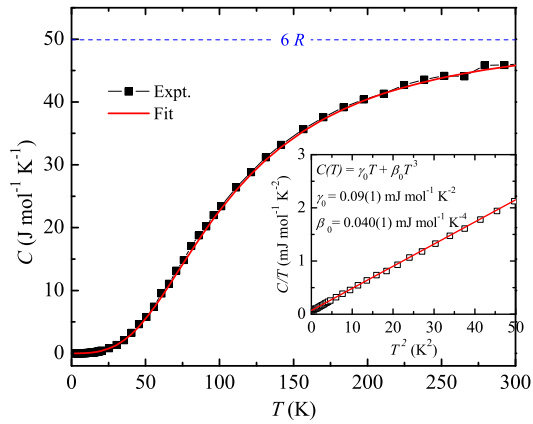}
\caption{\label{Fig.5} Main frame, temperature dependence of specific heat. The red curve is a Debye-model fit with $\Theta_D=$ 440 K. Inset shows $C/T$ vs. $T^2$ in the low temperature regime.}
\end{figure}

Fig.~\ref{Fig.5} shows the specific heat ($C$) as a function of temperature. For high temperature, $C(T)$ follows the Dulong-Petit law. At 300 K, $C$ reaches 46.07 J/(mol$\cdot$K), which is close to but somewhat smaller than the classical limit $6R=$~49.88 J/(mol$\cdot$K), where $R=$ 8.314 J/(mol$\cdot$K) is the ideal gas constant. Below 7 K, $C(T)$ well conforms to $C(T)=\gamma_0 T+\beta_0 T^3$, which is better expressed by the $C/T$ versus $T^2$ plot inset to Fig.~\ref{Fig.5}. The derived Sommerfeld coefficient is $\gamma_0=$ 0.09(1) mJ/(mol$\cdot$K$^2$). This vanishingly small $\gamma_0$ corroborates a very low electronic density of states, in agreement with the semimetallicity and low carrier density of NbAs. We also obtained $\beta_0=$ 0.040(1) mJ/(mol$\cdot$K$^4$). Such a small $\beta_0$ value is reminiscent of a high Debye temperature $\Theta_D$. By fitting the heat capacity data up to 300 K to the Debye model, we get $\Theta_D=$ 440 K. It should be noted that a direct estimate from $\beta_0$ yields $\Theta_D=(\frac{12\pi^4NR}{5\beta_0})^{1/3}=$ 458 K ($N=$ 2 for NbAs)\cite{Ashcroft-SSP}. The uncertainty places $\Theta_D$ in the vicinity of 450(9) K. This magnitude of $\Theta_D$ is much higher than that of the elementary constituents, Nb (285 K) and As (278 K)\cite{Ashcroft-SSP}. Since $\Theta_D$ is a measure of the maximum phonon frequency and thus the stability of the crystalline lattice, the large $\Theta_D$ in NbAs indicates a small compressibility, which accounts for the relatively weak pressure effect aforementioned. Recently, Zhou {\it et al.} found that TaAs, another member of the $TmPn$-family of Weyl semimetals, displays insulating-like behavior above 1.1 GPa and a probable structural transition near 10.0 GPa\cite{ZhouY-TaAsPre}. Such a structural transition is absent in NbAs for pressures up to 26.0 GPa\cite{Zhang-NbAsPre}. Apparently, the crystalline  lattice of TaAs is less stable than of NbAs. This might be not surprising considering that the $\Theta_D$ of a $5d$-transition-metal arsenide would be smaller than that of a $4d$-transition-metal arsenide. The large Nb-$4d$ bandwidth\cite{LeeC-TmPnband} and large Fermi energy also make the electronic state of NbAs robust against modest pressure. Whether metal-insulator and structural phase transitions will occur in NbAs under higher pressure is still an open question.


Interestingly, by examining the Kadowaki-Woods (KW) ratio $R_{KW}\equiv A/\gamma_0^2$\cite{Kadowaki-KW}, where $A$ is the prefactor of the $T^2$ term in $\rho_{xx}(T)$ , one can obtain new insight into the scattering rate in this Weyl semimetal. $\rho_{xx}(T)$ in NbAs obeys a $T^2$-power law from 1.3 K up to 20 K with $A=2.6(2)\times10^{-7}$ m$\Omega$ cm/K$^2$ as shown in the inset to Fig.~\ref{Fig.4}(c), although the inelastic contribution remains small relative to the elastic scattering as measured by the residual resistivity. The resulting KW ratio $R_{KW}=3.2\times10^{4}$ $\mu\Omega$ cm mol$^2$ K$^2$ J$^{-2}$ is significantly enhanced over the value of 10 $\mu\Omega$ cm mol$^2$ K$^2$ J$^{-2}$ found in HF systems\cite{Kadowaki-KW} and 0.4 $\mu\Omega$ cm mol$^2$ K$^2$ J$^{-2}$ in transition metals\cite{Rice-KW}. Jacko {\it et al.} explained that a large enhancement is expected in low-carrier-density systems\cite{Jacko-KW2009}. They proposed a modified KW ratio $R_{KW}^*\equiv Af(n)/\gamma_0^2$, where $f(n)$ is a universal function that depends on the carrier density $n$ and the dimensionality of the system. In the subsequent analysis we assume that the conductivity is dominated by the hole-like $\beta$-pocket since the mobility of these carriers is significantly larger than the electron-like $\alpha$-pocket. Using $f(n)=(3n^7/\pi^4\hbar^6)^{1/3}$ as expected for a 3D system (Note \cite{note}) we find $R_{KW}^*= 6.8 \times 10^{112}$ K$^2$ J$^{-3}$ s$^{-1}$ C$^{-2}$.
This is more than 5 orders of magnitude less than the expected value of 1.25$\times10^{118}$ K$^2$ J$^{-3}$ s$^{-1}$ C$^{-2}$ found to be in good agreement for HFs, transition metals, organics and transition-metal oxides alike\cite{Jacko-KW2009}. Clearly, the scattering rate is dramatically suppressed in comparison with the expectations based solely on carrier density and density of states.
It is rather natural to attribute this measure of suppressed scattering to the topological nature of the Weyl nodes which suppresses backscattering.
We also notice that the pressure dependence of the $A$ coefficient is significantly less than that of the residual resistivity. $A$ increases by less than 10\%, but the residual resistivity increases by more than 30\%. In other words, the inelastic scattering rate increases much more weakly than the elastic scattering rate under pressure. Finally, it might be of some interest to consider the $T^2$ coefficient of resistivity to have the general form of $A = \hbar/e^2 (k_B/\varepsilon_F)^2 l_{quad}$, where $l_{quad}$ is a material dependent quantity associated with microscopic details of momentum decay by scattering, as was done in a recent analysis of low-carrier-density SrTiO$_{3-\delta}$ samples\cite{Lin-T2Sci}. Using the $\varepsilon_F$ value of the $\beta$-pocket we find that $l_{quad} = 1.3$ nm, a value lying between 1 and 40 nm that covers a variety of materials\cite{Lin-T2Sci}. In particular, this value is very close to those of the low carrier density systems like bismuth and graphite. The origin of this commonality of $l_{quad}$ is an interesting problem for the future.

\section{Conclusion}

In conclusion, pressure dependent magnetotransport properties of the Weyl semimetal NbAs have been studied under hydrostatic pressure up to 2.31 GPa. $\rho_{xx}(T)$ curves change very little with pressure, and there is no signature of superconductivity down to 0.3 K. The Fermi surfaces (both $\alpha$- and $\beta$-pockets) undergo an anisotropic evolution with pressure: the extremal areas slightly increase in the $\mathbf{k_x}$-$\mathbf{k_y}$ plane, but decrease in the $\mathbf{k_z}$-$\mathbf{k_y}$($\mathbf{k_x}$) plane. The topological features of the two pockets remain unchanged. By fitting the temperature dependence of specific heat to the Debye model, we obtained a large Debye temperature for this compound, $\Theta_D=$ 450(9) K, indicative of a ``hard" crystalline lattice that corroborates this relatively weak pressure effect. External pressure may introduce a small degree of inhomogeneity or strain, and consequently the elastic scattering rate is enhanced. We also calculated the Kadowaki-Woods ratio $R_{KW}=$ 3.2$\times 10^4$ $\mu\Omega$ cm mol$^2$ K$^2$ J$^{-2}$, and discussed the suppressed inelastic scattering in the context of the topological nature of a Weyl semimetal.

\section*{Acknowledgments}

We thank Hongchul Choi, Jianxin Zhu, Yaomin Dai and Xiao Lin for helpful discussions. Samples were synthesized under the auspices of the Department of Energy, Office of Basic Energy Sciences, Division of Materials Science and Engineering. Electrical transport measurements were supported by the LANL LDRD program. Y. Luo acknowledges a Director's Postdoctoral Fellowship supported through the Los Alamos LDRD program.

\section*{References}



\providecommand{\newblock}{}


\pagebreak[4]

\begin{table}
\tabcolsep 0pt \caption{\label{Tab.1} Comparison of physical quantities between $p$=0 and 2.31 GPa.}
\vspace*{-12pt}
\begin{center}
\def\temptablewidth{1.0\columnwidth}
{\rule{\temptablewidth}{1pt}}
\begin{tabular*}{\temptablewidth}{@{\extracolsep{\fill}}ccccc}
Quantities                                               & $\alpha$, 0 GPa       & $\alpha$, 2.31 GPa      & $\beta$, 0 GPa         & $\beta$, 2.31 GPa \\\hline
$F^a$, $F^{a'}$ (T), $\mathbf{B}$$\parallel$$\mathbf{a}$ & 101.5(3), 106.0(4)    & 90.3(3), 97.0(3)        & 94.0(3), 97.0(5)       & 82.1(2), 85.1(3)      \\
$F^c$ (T), $\mathbf{B}$$\parallel$$\mathbf{c}$           & 20.9(2)               & 21.7(2)                 & 15.6(2)                & 17.0(2)               \\
$k_F^c$ (\AA$^{-1}$)                                     & 0.0252(1)             & 0.0257(1)               & 0.0218(1)              & 0.0227(1)             \\
Intercept                                                & 0.35(1)               & 0.32(1)                 & 0.12(1)                & 0.13(1)               \\
$\Phi_B$                                                 & 0                     & 0                       & $\pi$                  & $\pi$                 \\
$m^*$ ($m_0$)                                            & 0.046(3)              & 0.041(1)                & 0.029(1)               & 0.026(1)              \\
$v_F$ (10$^5$ m/s)                                       & 6.34(4)               & 7.25(6)                 & 8.70(6)                & 10.12(7)              \\
$\varepsilon_F$ (meV)$^\dag$                             & 52.6(5)               & 61.3(5)                 & $-$124.7(7)            & $-$151.5(8)           \\
$n$ (10$^{18}$ cm$^{-3}$)$^\ddag$                        & ($-$)10.7(2)          & ($-$)9.8(2)             & ($+$)9.4(2)            & ($+$)8.4(2)           \\
$\mu$ (10$^{5}$ cm$^2$/Vs)$^\ddag$                       & 1.9(2)                & 1.5(2)                  & 19(3)                  & 17(3)                 \\
$\tau_{tr}$ (10$^{-12}$ s)                               & 5.0(3)                & 3.5(2)                  & 31(2)                  & 27(2)                 \\
$l_{tr}$ ($\mu$m)                                        & 3.2(6)                & 2.5(5)                  & 27(2)                  & 25(2)                 \\
\end{tabular*}
{\rule{\temptablewidth}{1pt}}
\end{center}
\begin{flushleft}
$^\dag$ The Fermi energy for the $\alpha$-pocket is referenced from the bottom of the parabolic valence band (electron-type), while for the $\beta$-pocket it is referenced from the Weyl point (hole-type).\\
$^\ddag$ Estimated from the Hall effect.\\
\end{flushleft}
\end{table}

\newpage

\end{document}